\documentclass[aps,pre,reprint,amsmath,amssymb]{revtex4-1}
\usepackage{graphicx,subfigure}
\usepackage{dcolumn}
\usepackage{epstopdf}
\usepackage{upgreek}
\usepackage{textcomp}
\usepackage{ulem}
\usepackage{bm}
\usepackage{color}
\usepackage{xcolor}

\newcommand{\D}[1]{\textcolor{violet}{#1}}

\makeatletter
\def\btt#1{\texttt{\@backslashchar#1}}%
\DeclareRobustCommand\bblash{\btt{\@backslashchar}}%
\makeatother

\begin{document}
\title{Electric field driven controllable motility of metal-dielectric Janus particles with boojum defects in a nematic liquid crystal}

\date{\today}
\author{Dinesh Kumar Sahu, and Surajit Dhara$^{*}$}
\affiliation{School of Physics, University of Hyderabad, Hyderabad 500 046, India
}

\begin{abstract}
 
In a sharp contrast to the response of silica particles we show that the metal-dielectric Janus particles with boojum defects in a nematic liquid crystal are self-propelled under the action of an electric field applied perpendicular to the director. The particles can be transported along any direction in the plane of the sample by selecting the appropriate orientation of the Janus vector with respect to the director.  The direction of motion of the particles is controllable by varying the field amplitude and frequency. 
 The command demonstrated on the motility of the particles is promising for tunable transport and microrobotic applications.
 
\end{abstract}
\maketitle


Transport of tiny particles in fluids by electric field, the so-called electrophoresis \cite{ramos}, is an active area of research in soft matter science and widely used in segregating macromolecules \cite{drop}, colloidal-assembly and transport \cite{stv1,div2}, biomedical \cite{morgan,drop}, microfluidic \cite{alex} and electrophoretic display devices \cite{com,rc}. Active control and piloting of the particles in the suspension is of crucial importance for developing new technologies. In linear electrophoresis, the ions in the electrical double layer drag the fluid (electro-osmosis) as a result the freely suspended charged particles move along the applied field direction with a velocity proportional to the field (${\bf E}$). In nonlinear electrophoresis, known as induced-charge electroosmosis (ICEO), the applied field creates a double layer and drags the fluid, setting a flow pattern of quadrupolar symmetry surrounding the particles \cite{tm1,baz1,baz2,baz3,tm}. The quadrupolar symmetry of the ICEO flows is broken for polarisable particles with nonuniform surface properties or polarity in the shape, consequently the particles move with a velocity proportional to $E^2$\cite{baz3}. However, the transport of particles by the electric field in either type is unidirectional, without having any navigational capabilities \cite{sum}.

\begin{figure}[!ht]
\centering
\includegraphics[scale=0.47]{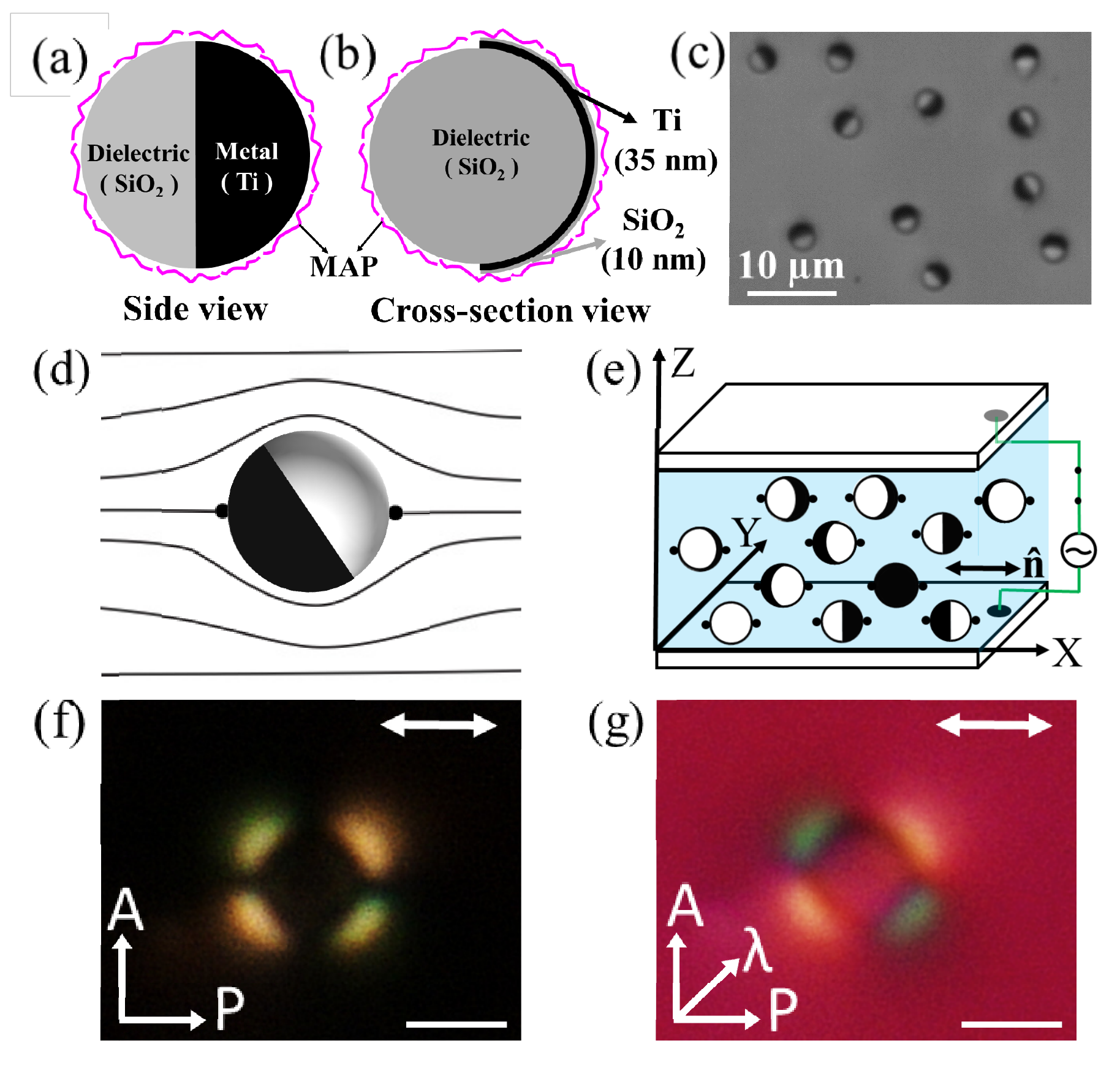}

\caption{(a,b) Schematic diagram of a metal-dielectric Janus particle. MAP coating is indicated by magenta strands. (c) Optical micrograph of the Janus particles. Darker (brighter) region represents the metal (dielectric). (d) Director field surrounding a Janus particle with Boojum defects. (e) Diagram of a cell made of parallel electrodes. (f) Polarising optical micrograph of a Janus Boojum-particle. (g) Micrograph with a full-wave plate (530 nm). Double headed arrows represent director $\bf{\hat{n}}$. Scale bar: 3 $\upmu$m.}   
\label{fig:figure1}
\end{figure}

\begin{figure*}[!ht]
\begin{center}
\includegraphics[scale=0.7]{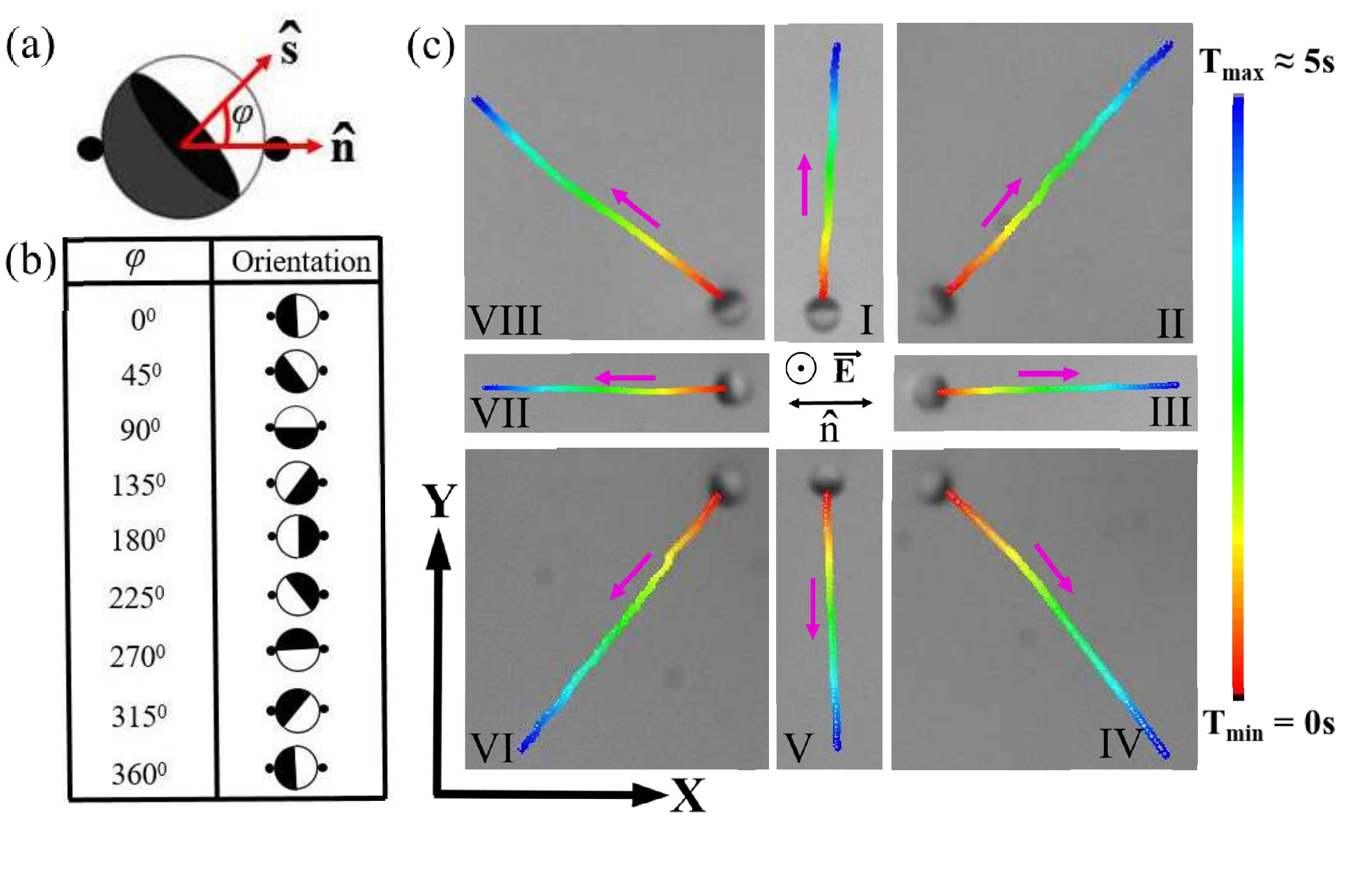}
\end{center}
\caption{ (a) Janus vector ${\bf \hat s}$ is normal to the metal-dielectric interface. $\varphi$ is the angle between ${\bf \hat s}$ and the far field director ${\bf \hat n}$. (b) Particles with selected $\varphi$ chosen for the experiments. (c) Time coded trajectories in the $xy$-plane, labelled from I to VIII, under ac electric field (2.38 V$\upmu $m$^{-1}$, 30 Hz). Trajectories of selected particles are grouped.  Pink arrows denote the direction of motions. (Movie S1 ~\cite{sup}). The direction of electric field {\bf E} (along $z$-direction) and the director ${\bf \hat n}$ for all the trajectories are shown at the centre. (T\textsubscript{min}=0 s to T\textsubscript{max}=5 s)}
\label{fig:figure2}
\end{figure*}

The origin of electrophoretic motion of colloidal particles in nematic liquid crystals (NLCs) is quite different from that of the isotropic counter part \cite{od,od1,oleg,oleg1,oleg2,id}. In NLCs, the particles with perpendicular anchoring of director $\hat{\bf n}$ (the average direction of molecular orientation) create elastic deformation in the medium, inducing point \cite{lub,stark} or equatorial disclination ring defects (so-called Saturn ring) \cite{ram,abot}. Such particles are known as elastic dipoles and quadrupoles, respectively \cite{stark}. When the director anchoring is planar, two antipodal surface defects (boojums) are induced along the far-field director~\cite{zuhail}. The particles with dipolar director configuration exhibit bidirectional transport in which the fore-aft symmetry of the electro-osmotic flows is broken due to the point defect~\cite{od,od1}. Electrophoresis of such defect-decorated particles is nonlinear and known as LC-enabled electrophoresis (LCEEP)\cite{oleg}. The particles with quadrupolar director configuration inducing symmetric defects such as Saturn ring or Boojums are immobile as the fore-aft symmetry of the electroosmotic flow is retained~\cite{oleg2}. Very recently Sahu \textit{et al.} have shown that Janus particles decorated with Saturn-ring defect can be transported and navigated at will by external ac electric fields~\cite{sd}. Here, we study the effect of ac electric field on Janus particles with planar anchoring of the LC director that induces two antipodal surface defects. We show that the Janus Boojum-particles become motile and the direction of motion depends on the relative orientation of the metal-dielectric interface with respect to the far-field director $\hat{\bf n}$. Further, the direction of motion of the particles is controllable by changing the frequency and the amplitude of the ac field. 


The metal-dielectric Janus particles are synthesized by adapting the following the procedure. To begin with, an aqueous suspension (~2$\%$) of silica particles (SiO\textsubscript{2}) of diameter $2a = 3.0 \pm 0.2$ $\upmu$m (Bangs Laboratories, USA) are spread over a glass slide treated with Piranha solution to form a colloidal monolayer. 
A layer of Titanium (Ti), thickness approximately 35 nm is deposited on the top of the colloidal monolayer using an electron-beam deposition technique so that only the exposed upper-hemispheres of the particles get coated (Fig.\ref{fig:figure1}(a)). Further, a thinner layer (10 nm) of SiO\textsubscript{2} is deposited to make the overall particle's surface chemically isotropic (Fig.\ref{fig:figure1}(b)). Finally the particles are extracted from the slides through ultrasonication in distilled water. An optical micrograph of a few Janus particles is shown in Fig.\ref{fig:figure1}(c). The metal hemisphere appears dark under transmitted light.

 Next the Janus particles are coated with N-Methyl-3 aminopropyl trimethoxysilane (MAP) to induce planar or homogeneous anchoring of the LC molecules (Fig.\ref{fig:figure1}(a,b)). A small quantity of Janus particles (0.01 wt\%) is dispersed in the nematic liquid crystal (MLC-6608, Merck) that exhibits the following phase transitions: smectic-A -30\textsuperscript{o}C nematic 90\textsuperscript{o}C Isotropic. At room temperature the dielectric anisotropy is negative ($\Delta\epsilon=\epsilon_{\parallel}-\epsilon_{\perp}$=-3.3) and and the conductivity anisotropy is positive ($\Delta\sigma=\sigma_{||}-\sigma_{\perp}\simeq6.1\times10^{-10}$ S/m)~\cite{pra}. The subscripts $\parallel$ and $\perp$ denote the quantities measured parallel and perpendicular to the director $\bf\hat{n}$.
The colloidal mixtures are studied in cells made of two indium-tin-oxide (ITO) coated glass plates (Fig.\ref{fig:figure1}(e)). The ITO coated glass plates are pretreated with a polyimide AL-1254 (JSR Corporation, Japan) and cured at 180$^{\circ}$C for 1 hour and rubbed unidirectionally using a bench top rubbing machine (HO-IAD-BTR-01) for planar alignment of the  director. Typical cell thickness used in the experiment is in the range of 5-7 $\upmu$m.\\
We used an inverted polarising optical microscope (Nikon Ti-U) with water immersion objective (Nikon, NIR Apo 60/1.0) for the experiments. For particle manipulation, a dynamic laser tweezer with a cw solid-state laser operating at 1064 nm (Aresis, Tweez 250si) was used. A camera (iDs-UI) attached to the microscope is used for  recording the particle trajectories at the frame rate of 50-60 per second. The position of the particle is tracked off-line using a computer program with an accuracy of $\pm$10 nm.\\
  The MAP coated Janus particles in nematic LC induce two antipodal point defects (Boojums) as shown in Fig.\ref{fig:figure1}(d). Figure \ref{fig:figure1}(f) shows the polarising optical micrograph of a Janus Boojum-particle in the nematic LC (MLC-6608). The surface anchoring of the LC molecules and the resulting elastic deformation surrounding the Janus particle are confirmed from the texture obtained by inserting a full-wave plate or $\lambda$-plate (530 nm) as shown in Fig.\ref{fig:figure1}(g). The
magenta colour corresponds to the parallel orientation of the long axis of the molecules in the rubbing direction (e.g., the far field director ${\bf \hat n}$), whereas the bluish and yellowish colours surrounding the particle correspond to the anticlock-wise and clockwise rotation of the director with respect to the rubbing direction~\cite{zuhail}.\\

  When no electric field is applied the Janus vectors ${\bf \hat s}$  of the particles (normal to the plane of the metal-dielectric interface, see Fig.\ref{fig:figure2}(a))  are oriented in arbitrary directions. 
  When the electric field is switched on the particles reorient such that the metal-dielectric interface becomes parallel to {\bf E} to minimise the torque due to the induced dipole moment~\cite{sd,pra}. The applied field, however does not influence the macroscopic director except near the particles as the dielectric anisotropy of the sample is negative~\cite{oleg1}. In Fig.\ref{fig:figure2}(b) we present orientation of the Janus vectors  of a few selected particles, which we investigated. 
 As the field is increased beyond a particular value the particles start moving along some specific directions in the plane of the sample, depending on the orientation of the Janus vector ${\bf\hat{s}}$ with respect to the director. Real-time trajectories of the particles at a fixed field are grouped (I-VIII) and shown in Fig.\ref{fig:figure2}(c). It is noted that for all trajectories the particles are moving, facing the the metal hemisphere [Movie S1~\cite{sup}]. This is in contrast to the motion of the quadrupolar particles in which the motion is not restricted to metal facing. With the help of the laser tweezer, the angle $\varphi$ between the Janus vector ${\bf\hat{s}}$ and the director $\bf\hat{n}$ can be changed by photothermal quenching of the surrounding director field and the particle can thus be transported in any predetermined direction as desired.
 
 \begin{figure}
\centering
\includegraphics[scale=0.25]{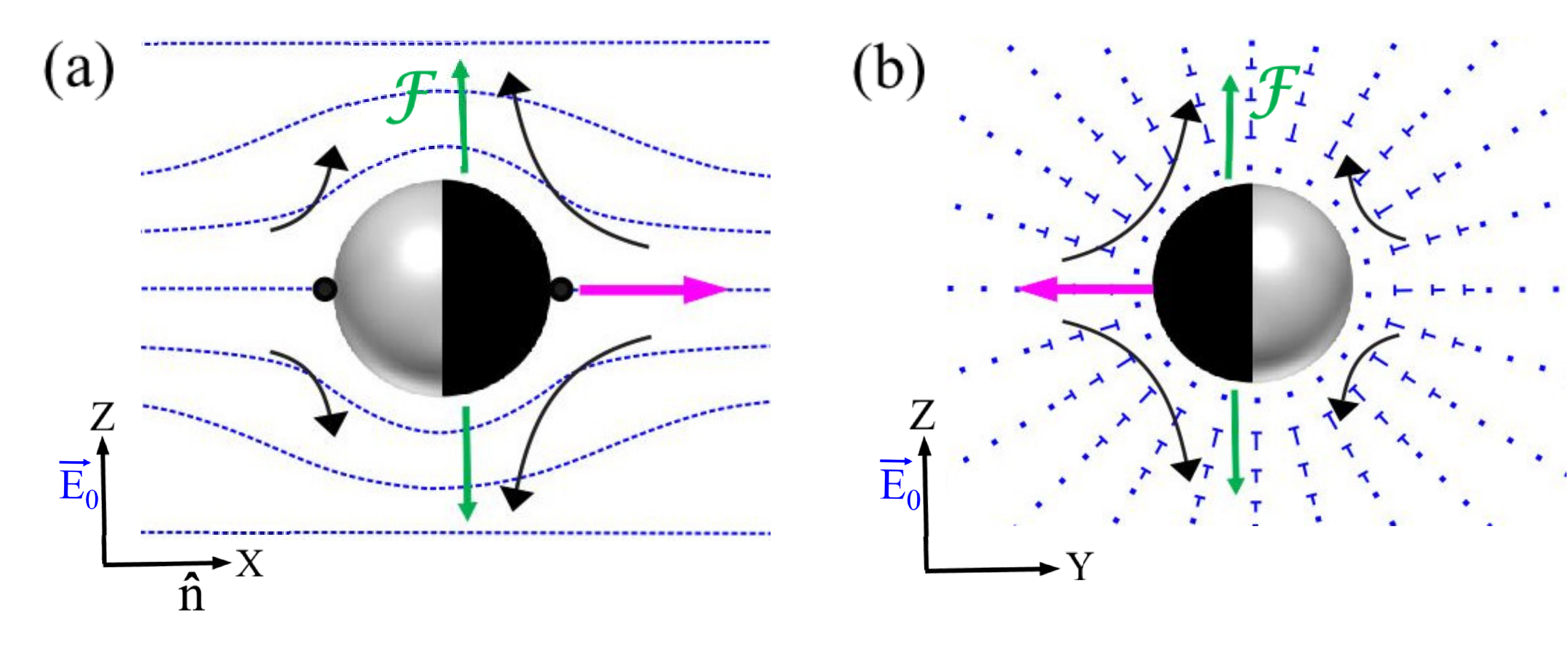}
\caption{ Director field and proposed electroosmotic flows surrounding a Janus Boojum-particle in two orthogonal planes parallel to the electric field {\bf E}. The large curved arrows on the metal hemisphere indicate stronger flows. The off-centered force dipoles $\boldsymbol{\mathcal{F}}$ (shown in green colour) are pushers in both the planes. The propulsion directions of the particles are indicated by pink arrow.}
\label{fig:figure3}
\end{figure}
 
 The motility of the Janus particles with Saturn-ring defect was accounted for theoretically from the electrostatic force density induced by the field which drags the fluid in the vicinity of the particles~\cite{sd}. The force depends on the field strength, dielectric anisotropy, conductivity anisotropy and the curvature of the director around the particles. Within a linearised theory, for particles with homeotropic boundary condition and assuming the deviation of director from the mean alignment is small, it was
shown that the resulting force dipole was either a puller or a pusher depending on the orientation of the Janus vector. However, for particles with planar anchoring the director deviation near the defects is not small and the force density equation can not be simplified analytically. A numerical solution of the complete nonlinear theory will be required for getting quantitative values of the force dipoles, which is beyond the scope of the present work. It is observed in Fig.\ref{fig:figure2}(c) that the particles move with the metal hemisphere forward irrespective of the orientation of the Janus vector in relation to the nematic axis. We conjecture that the effective force dipole is pusher type with respect to the direction of the electric field. Based on this assumption, we propose possible flow directions in two orthogonal planes as shown in Fig \ref{fig:figure3}. Since the metal hemisphere is highly polarised the flow is stronger on that side relative to the dielectric hemisphere resulting in an off-centred force dipole as shown by the green arrows in Fig \ref{fig:figure3}.

\begin{figure}[!h]
\centering
\includegraphics[scale=0.45]{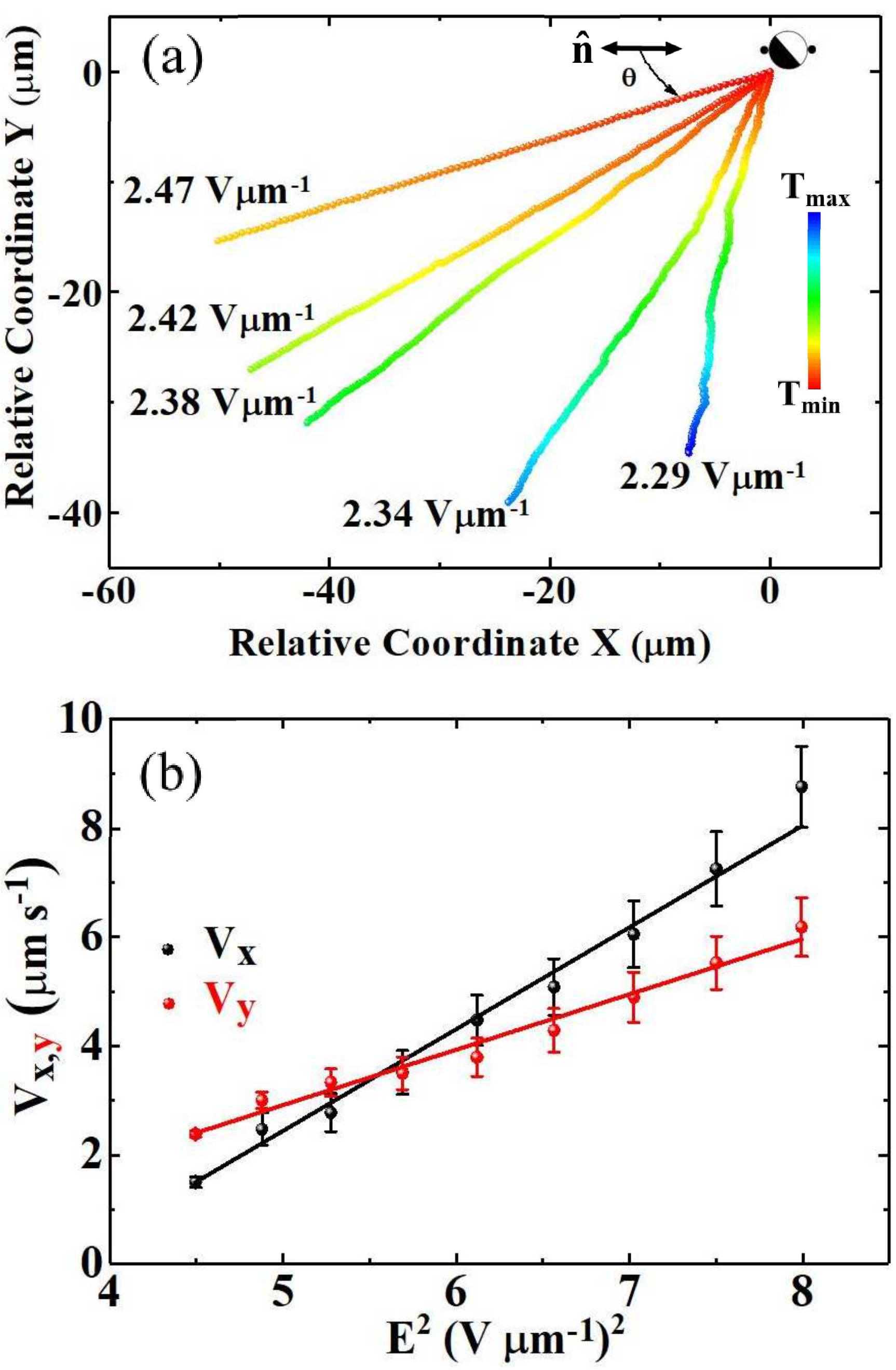}
\caption{(a) Time-coded trajectories of Janus particles at different field amplitudes and at a fixed frequency (30 Hz) for a fixed orientation ($\varphi=45^{\circ}$) of the Janus vector. T\textsubscript{max}$= 14.5$ s and T\textsubscript{min}$= 0$ s. $\theta$ is the angle of the trajectories with respect to the director. (b) Velocity components $V_x$ and $V_y$ at different fields obtained from (a). Slopes of $V_x$ and $V_y$ are 1.85 and 1.01 $\upmu$m\textsuperscript{3} s\textsuperscript{-1} V\textsuperscript{-2} respectively.  }
\label{fig:figure4}
\end{figure}

\begin{figure}[!h]
\centering
\includegraphics[scale=0.5]{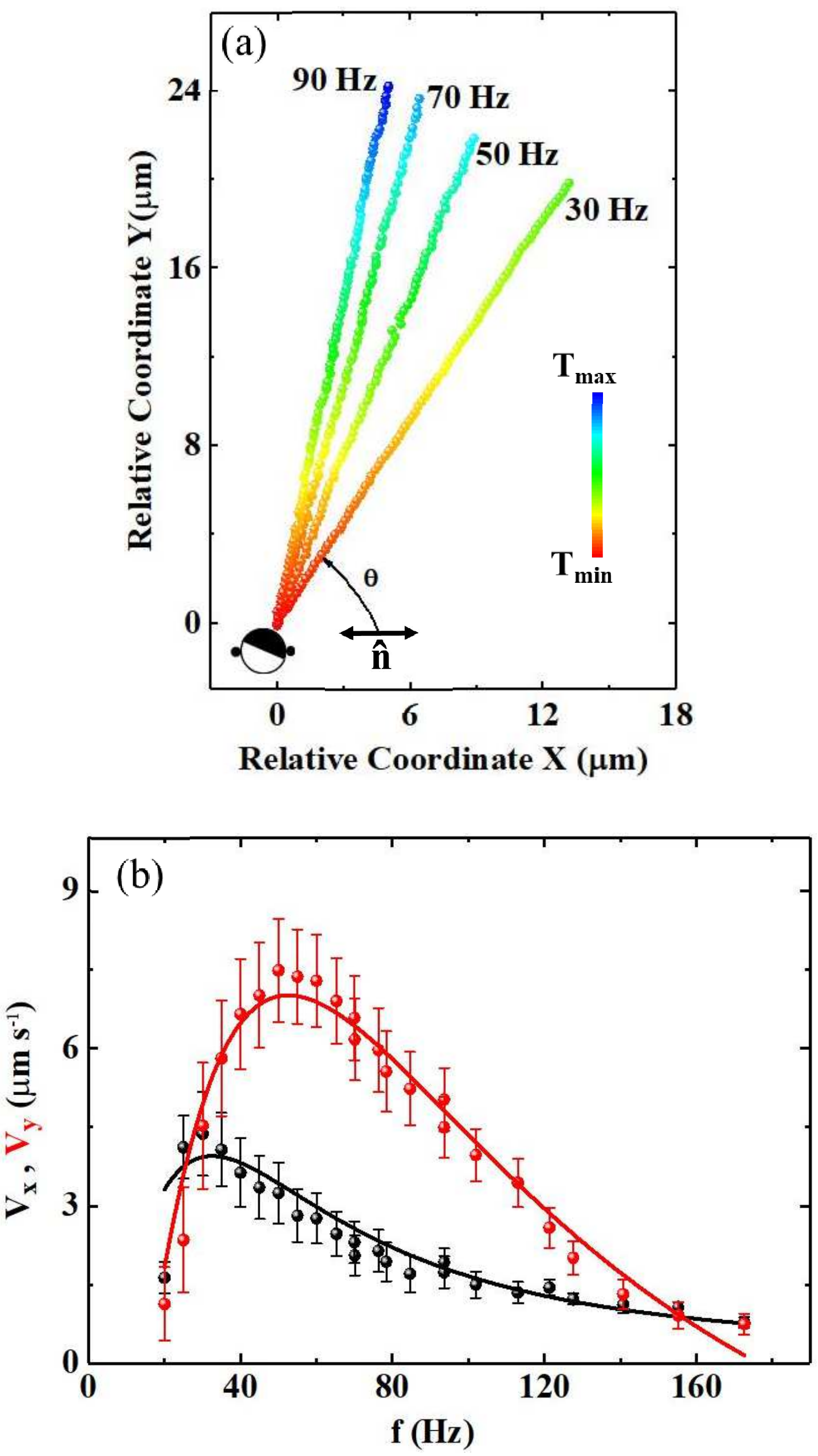}
\caption{(a) Time coded trajectories of a particle at different frequencies and at a fixed amplitude (2.38 V$\upmu$m$^{-1}$) for a fixed orientation ($\varphi=250^{\circ}$) of the Janus vector. T\textsubscript{max}$=10$ s and T\textsubscript{min}$=0$ s. $\theta$ is the angle between the far field director and the trajectory. (b) Velocity components $V_x$ and $V_y$ at different frequencies at $E=2.38$ V$ \upmu $m$^{-1}$. Red and black lines show theoretical fit to equation (1). Fit parameters are: $\tau_{e}= 0.05$ s, $\tau_{p}= 0.014$ s for $V_{x}$ and $\tau_{e}= 0.045$ s, $\tau_{p}= 0.008$ s for $V_{y}$. Error bars represent the standard deviation of the mean value. }
\label{fig:figure5}
\end{figure}

For a fixed orientation of the Janus vector the trajectory also depends on the amplitude and frequency of the applied electric field.
Figure \ref{fig:figure4}(a) shows that for a fixed orientation of the Janus vector, the trajectory depends on the amplitude of the field. In particular, for the Janus vector $\varphi=45^{\circ}$, the angle ($\theta$) between the trajectory and $\bf{\hat{n}}$ at a fixed frequency decreases with increasing field amplitude from 2.29 V$\upmu$m$^{-1}$ to 2.47 V$\upmu$m$^{-1}$. This effect can be explained from the relative variation of the velocity components  $V_x$ and $V_y$ as shown in Fig.\ref{fig:figure4}(b). As expected both $V_x$ and $V_y$ increases linearly with the squared of the field but $V_x$ increases at a much faster rate than $V_y$. When the field amplitude is increased, the relative enhancements of the velocity components are unequal, namely $\Delta V_x$ is larger than $\Delta V_y$, and as a result the angle between the director $\bf{\hat{n}}$ and the trajectory, $\theta=\tan^{-1}(\Delta V_y/\Delta V_x)$ decreases.
 
 For a particle with fixed orientation of the Janus vector the trajectory also depends on the frequency as shown in Fig.\ref{fig:figure5}(a). For example, when the frequency of the field is increased from 30 to 90 Hz, the angle between the trajectory of the particle and the director ($\theta$) is increased. This effect can be explained from the relative variation of the velocity components  $V_x$ and $V_y$ with frequency ($f$) as shown in Fig.\ref{fig:figure5}(b). The motion of the particles is limited within a narrow range of frequency (30 - 170 Hz). The frequency dependence of both the components are fitted to the following equation~\cite{baz1,oleg1}
\begin{equation}
V_{i}(\omega) = {V^{o}_{i}} \frac{\omega^{2} \tau^{2}_{e}}{(1+\omega^{2} \tau^{2}_{p})(1+\omega^{2}\tau^{2}_{e})} 
\end{equation}
where $i=x,y$; $\omega=2\pi f$ is the angular frequency of the applied field, $\tau_{e}$ and $\tau_{p}$ are the characteristic electrode and particle charging time, respectively. In the low frequency region both the components increase as $f^2$ but decrease as $1/f^2$ in the high frequency region. Beyond the peak frequency (47Hz), the velocity decreases as $1/f^2$ but the coefficient of decrease of $V_x$ is larger than that of $V_y$ as a result the angle $\theta$ increases with the increasing frequency. Similar effect was observed in case of quadrupolar particles at slightly different field amplitude and frequency~\cite{sd}.
 
 \begin{figure}[!h]
\centering
\includegraphics[scale=0.55]{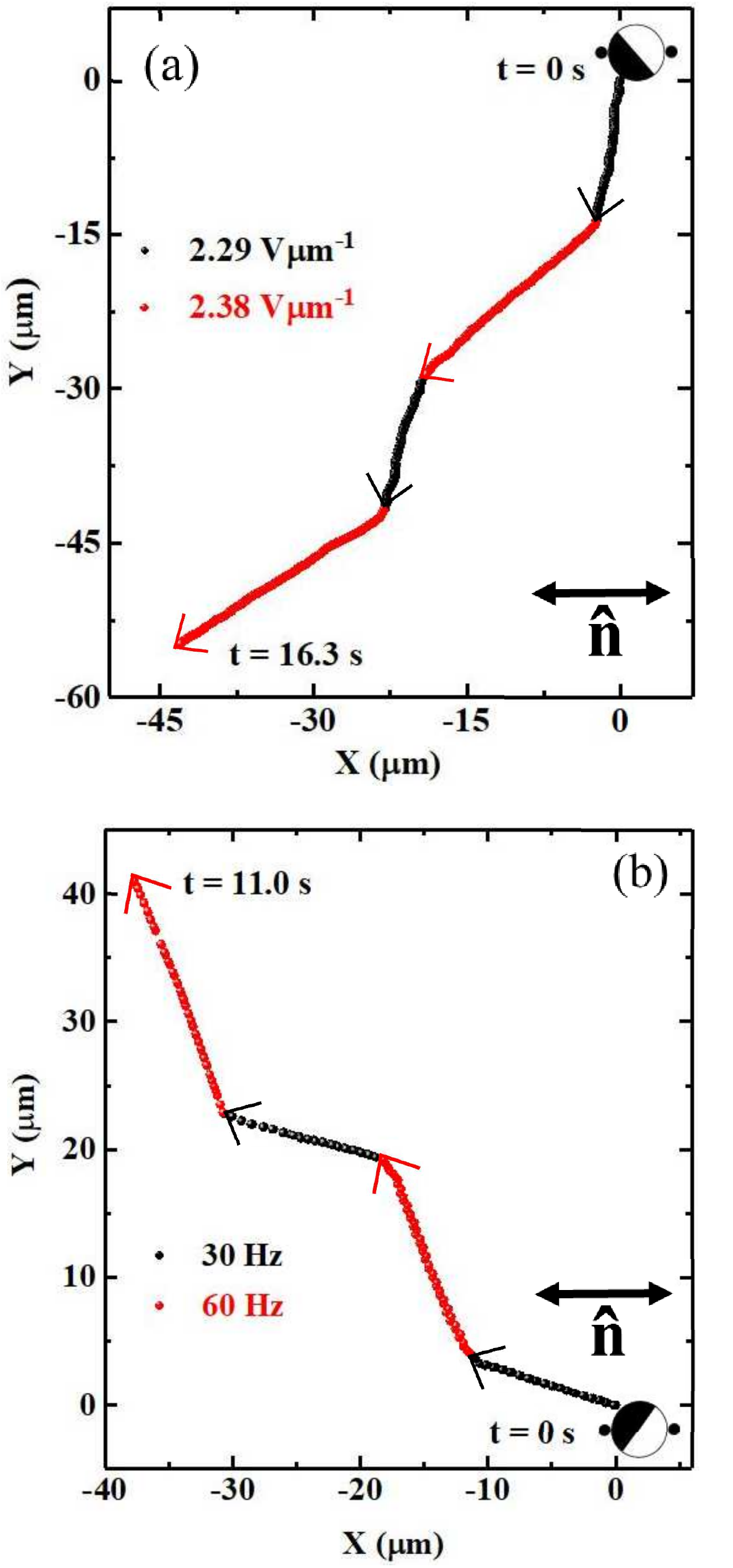}
\caption{(a) Change of direction of motion of a Janus particle with a fixed orientation ($\varphi=45^{\circ}$) by altering the field amplitude recursively between 2.29 V$\upmu$m$^{-1}$ (black) to 2.38 V$\upmu$m$^{-1}$ (red), keeping the frequency constant at 30 Hz (see Movie S2~\cite{sup}). (b) Change of direction of motion of a janus particle ($\varphi=315^{\circ}$) by altering the frequency recursively between 30 Hz (Black) to 60 Hz (Red) at a fixed field 2.38 V$\upmu$m$^{-1}$. (see Movie S3~\cite{sup}). }
\label{fig:figure6}
\end{figure}

 The effect of the changing field amplitude and frequency is exploited for navigating the direction of motion of the particles. Figure \ref{fig:figure6}(a) demonstrates the command over the trajectory by varying the  field amplitude recursively between two slightly different amplitudes, namely 2.29 and 2.38 V$\upmu$m$^{-1}$. For example, when the field amplitude is increased from 2.29 to 2.38 V$\upmu$m$^{-1}$, the particle changes it's direction of motion with respect to the initial direction by nearly 37$^{\circ}$ (Fig.\ref{fig:figure6}(a)). Figure \ref{fig:figure6}(b) demonstrates the navigation capability by changing the frequency recursively between 30 and 60 Hz. For example, the change of angle in it's direction of motion due to the change of frequency from 30 to 60 Hz is about 50$^{\circ}$. Thus, by changing either the amplitude or the frequency of the field or both, we can navigate the particles as desired. 
It may be mentioned that the effect of inertia is negligibly small and the director configuration around the particle remains unaffected as the Reynolds and the Ericksen numbers both are much less than 1~\cite{sd}.\\\\\\
   In conclusion we have studied the effect of ac electric field on metal-dielectric Janus particles inducing boojum defects. Although the director deformation induced by the particles is symmetric, the surface asymmetry of the particles break the fore-aft symmetry of the surrounding electroosmotic flows facilitating their motility. In all the trajectories the particles move facing the metal hemisphere and their direction of motion is dictated by the relative orientation of the Janus vector.  The direction of motion of the particles is manoeuvrable by changing the amplitude and frequency of the field. Such a spectacular control originates from the relative sensitivity of the electroosmotic velocities to the asymmetric surface and local director structure. Our work deals with the spherical particles however there are several new shape asymmetric particles whose Janus character under applied field is unexplored.
   Further our study is limited to dilute concentration of particles. It is expected that studies on the motility at higher concentration and their collective dynamics will be interesting~\cite{saha1,saha2}. 

{\bf ACKNOWLEDGEMENTS:}
 S.D. thanks Steve Granick for hosting his visit to IBS, UNIST which resulted in very useful discussions.  We thank K.V. Raman for help in preparing Janus particles This work is supported by the DST, Govt. of India (DST/SJF/PSA-02/2014-2015). S.D. acknowledges a  Swarnajayanti Fellowship and D.K.S. an INSPIRE Fellowship from the DST. We thank Swapnil Kole and Sriram Ramaswamy from Indian Institute of Science, Bangalore for very useful discussion.

\end{document}